\documentclass[aip,apl,graphicx,reprint,longbibliography]{revtex4-1}
\usepackage{graphicx}%
\usepackage{amsmath,amssymb}%
\usepackage{xcolor}
\usepackage[colorlinks=true]{hyperref}
\hypersetup{%
	colorlinks,
	plainpages=false,
	pdfpagelabels,
	breaklinks,
	pdfview=Fit,
	bookmarksopen,
	bookmarksnumbered,
	linkcolor=blue,
	anchorcolor=blue!70!white,
	citecolor=blue!80!white,
	filecolor=black,
	urlcolor=red!50!black
}%
\usepackage[utf8]{inputenc}
\usepackage[T1]{fontenc}
\usepackage{mathptmx}

\begin{document}


\title{Fluorescence coherence with anodic aluminum oxide hybrid photonic-plasmonic structure} 



\author{A.R. Hashemi}
\email[]{arezahashemi@gmail.com}
\author{Mahmood Hosseini-Farzad}
\email[]{hosseinif@shirazu.ac.ir}
\affiliation{Department of Physics, College of Sciences, Shiraz University, Shiraz 71946-84795, Iran}


\date{\today}

\begin{abstract}
\begin{minipage}[t]{0.48\linewidth}
	A class of hybrid photonic-plasmonic structures (HPPS) with vertical cylindrical cavities is proposed and its performance in providing a coherent light from spontaneous emission is investigated. It is shown that the proposed easy-to-fabricate and robust anodic aluminum oxide structure dramatically enhances the temporal and spatial coherence compared to the previously examined HPPS. The physical mechanism of achieving the temporal and spatial coherence is explained based on the special optical properties of the proposed HPPS and it is discussed how this structure enables one adjusting hybrid photonic-plasmonic optical modes to obtain coherence for emitters of different frequencies.
\end{minipage}
\vtop{%
	\vskip-1ex
	\hbox{%
		\includegraphics[width=0.35\linewidth]{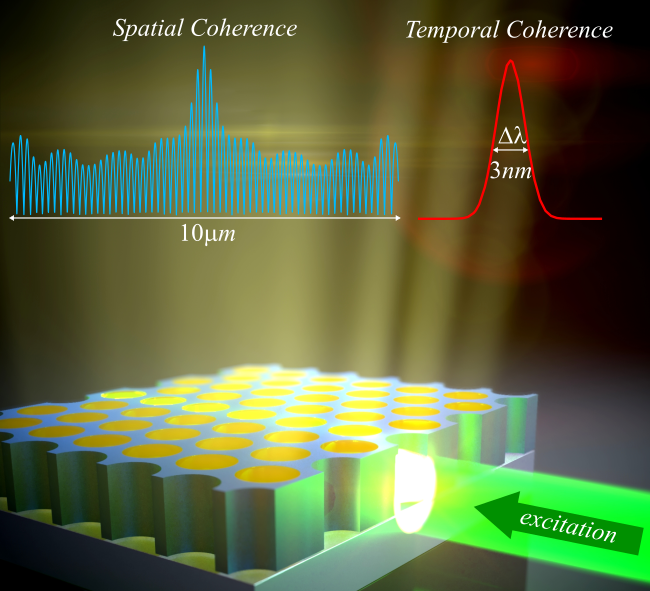}%
	}%
}%
\end{abstract}


\maketitle 

Fluorescence is a vastly used optical method for chemical and biological detection~\cite{Shcheslavskiy:18,giljohann2009}; however, its performance and efficiency shall be enhanced especially in sensing and imaging applications~\cite{ribeiro2017artefact,moerland2013shaping,zhao2014gold,wang2015plasmon}.	
Plasmonics appears to be the best method for tailoring and enhancing the fluorescence emission~\cite{Stockman2018,bauch2014}. It is known that light can be confined in close vicinity of metallic surfaces or nano-particles due to the coupling between electromagnetic (EM) waves and oscillations of electrical charges at the surface. The idea is to employ This interaction, which is referred to as surface plasmon resonance (SPR) in a way that a large enhancement in the optical density of states can be obtained in the neighborhood of fluorophores. This will be an effective means for elevating the excitation rate and raising the quantum yield as well as controlling the angular distribution of the fluorescence emission~\cite{kwon2008surface,kinkhabwala2009large,aouani2011plasmonic,lozano2013plasmonics,langguth2013plasmonic}.

Although the presence of fluorophore in the vicinity of metallic structures leads to the above mentioned appealing features, this adjacency may also increase the probability of quenching and non-radiative energy loss for the excited fluorophore~\cite{anger2006enhancement,pons2007quenching,li2009fluorescence,reineck2013distance}. This inadequacy can be resolved by means of a hybrid photonic-plasmonic structure (HPPS), that is created by attaching a photonic crystal (PC) to the metal surface (plasmonic structure). In other words, coupling between the surface plasmon polaritons (SPPs), i.e. the EM waves which are confined along the metal-dielectric interface, and the guided or trapped modes of the photonic crystal leads to the striking features of increasing the propagation length of SPPs, confining light in a deep subwavelength scale, and highly guided modes and cavity resonances in the PC structure~\cite{romanov2011,yang2011,zhang2012,schokker2017}. Moreover, by using an HPPS, the effective length of the evanescent normal component of SPPs can extend to tens of nanometers above the metal surface, therefore, the enhancement and directionality are obtained without any significant quenching~\cite{zhu2012broadband,lopez2010,ding2013spectral}.	

On the other hand, due to the weak correlation between the spontaneous emissions of fluorophores, the resulted light is isotropic in space and broad in spectrum, which in turn lower the detectability. Therefore, it is potentially advantageous to utilize a technique to create a coherent light from such spontaneous emitters. It must be noted that a desirable technique shall also prevent any significant loss in the emission intensity~\cite{raghunathan2012,greffet2002,de2012conversion}. In order to address these requirements, Shi and his coworkers~\cite{shi2014spatial,shi2014coherent} have proposed using a HPPS whose optical attributions was previously investigated by the group~\cite{shi2010optical}. It is shown that the interaction between the leaky modes (that can scape from propagating along the metal-dielectric interface) of HPPS and the fluorescent molecules successfully provides both the temporal and spatial coherence for the fluorescent emission. Such a technique can play an important role in different fluorescence applications. However, further investigations are needed to obtain a more efficient structure with an inherent capability to be adjusted for different spontaneous emitters. This is the aim of the present work.

Among different options for the photonic-crystal part of an HPPS, anodic aluminium oxide (AAO) presents a special feature; the vertical cylindrical cavities in the structure facilitates one's control over the direction of the propagation of the leaky modes. This provides the directionality of the emitted light and therefore can enhance the spatial coherence. Moreover, the frequency of the leaky modes can be adjusted by adapting the geometry of the cavities. 
In the literature, the capability of AAO in enhancing the intensity of fluorescent emission has also been addressed~\cite{li2012aluminum}.
Nevertheless, AAO has not yet been used in forming an HPPS to achieve coherent fluorescent emission.
In this work, a robust and easy to fabricate HPPS using the AAO structure is proposed that substantially enhances the coherence, while brings the required flexibility to be adjusted for spontaneous emitters with different excitation/emission frequencies.

\begin{figure}[t]
	\hspace{5ex}\includegraphics[width=0.99\columnwidth]{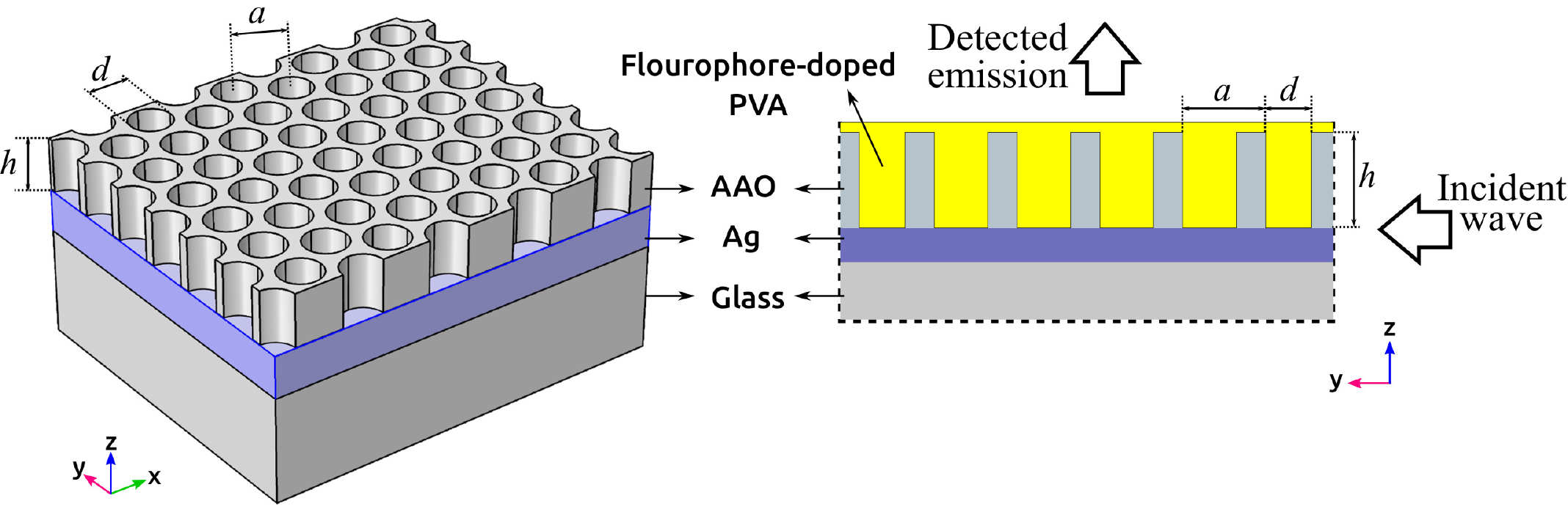}\\		
	\caption{\label{fig:AAOstr} The overall 3D view of the proposed hybrid anodic structure (left) and its $yz$ cross-section (right).}
\end{figure}

Here, the previously proposed method~\cite{ARXIV} which empowers finite difference time domain to simulate fluorescent molecules is applied to a novel HPPS that is constructed by placing an inverse photonic crystal, made by pore-opened\cite{Bruschi2015} anodic aluminium oxide (AAO), on top of a 200 nm thick silver (Ag) layer (Fig.\ref{fig:AAOstr}). The cylindrical holes of PC are filled with S101-doped PVA, which also forms a layer of 50 nm thickness on top of the AAO. The structural parameters, i.e., thickness of AAO layer $h=500$ nm, diameter of cylindrical holes $d=200$ nm, and the center-to-center distance the neighboring cylinders $a=250$ nm, have been set in a way that not only the structure can be easily fabricated, but also, it significantly enhances the coherence as seen in the rest of this paper.

\begin{figure}[t]
	\includegraphics[width=0.99\columnwidth]{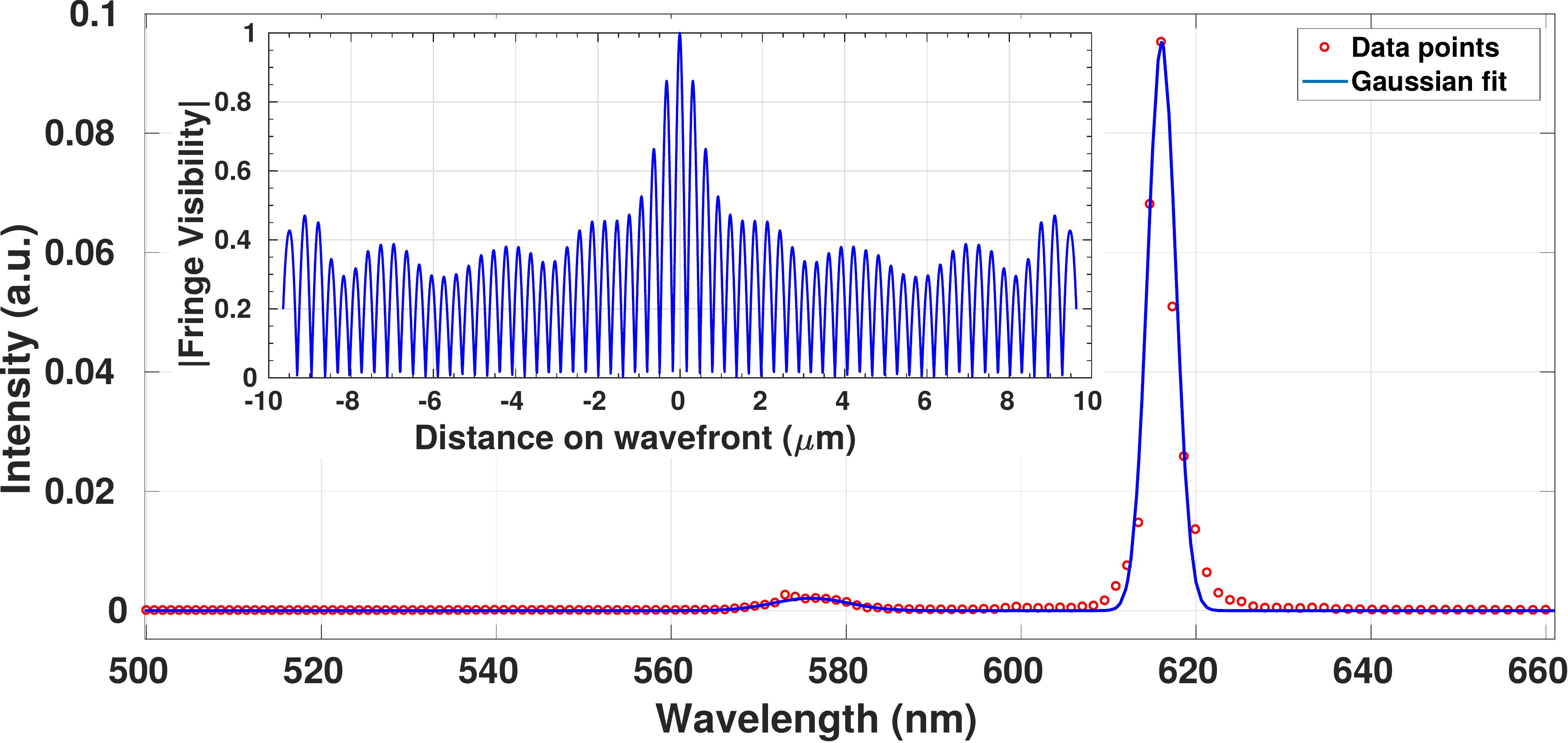}
	\caption{\label{fig:AAOFlu}Wavelength spectrum of the detected wave obtained for the proposed HPPS. 
		The emission pick is observed at $\lambda=616$ nm, while a lower pick occurs at  $\lambda=575$ nm that corresponds to the unabsorbed portion of the excitation wave.
		Shown in the inset is the fringe visibility as a function of the separation distance of the double-slits.}
\end{figure}

Similar to the previous test-case, the proposed HPPS experiences an incident continuous EM wave with a wavelength of 575 nm from its side that is perpendicular to the $y$-axis (Fig.~\ref{fig:AAOstr}) and the resulting vertical emission (along $z$-axis) is recorded.
In this way, the detected wave would not be masked by the incident wave. Moreover, this is a wise choice of the excitation and detection, which is appropriate for imaging and LED applications.   
In Fig.~\ref{fig:AAOFlu}, the recorded field is shown in frequency domain.
The graph hits its peak at $\lambda=616$ nm with a FWHM of $|\Delta \lambda| = 3$~nm. 
On the other hand, using TCF, a coherence time of $\tau_c=2.1\times10^{-13}$ sec or equivalently $|\Delta \lambda| = 4$~nm is calculated.
It is observed that the presence of the proposed HPPS leads to an almost eight times greater coherence length for the fluorescence emission. 

\begin{figure}[t]
	\includegraphics[width=0.99\columnwidth]{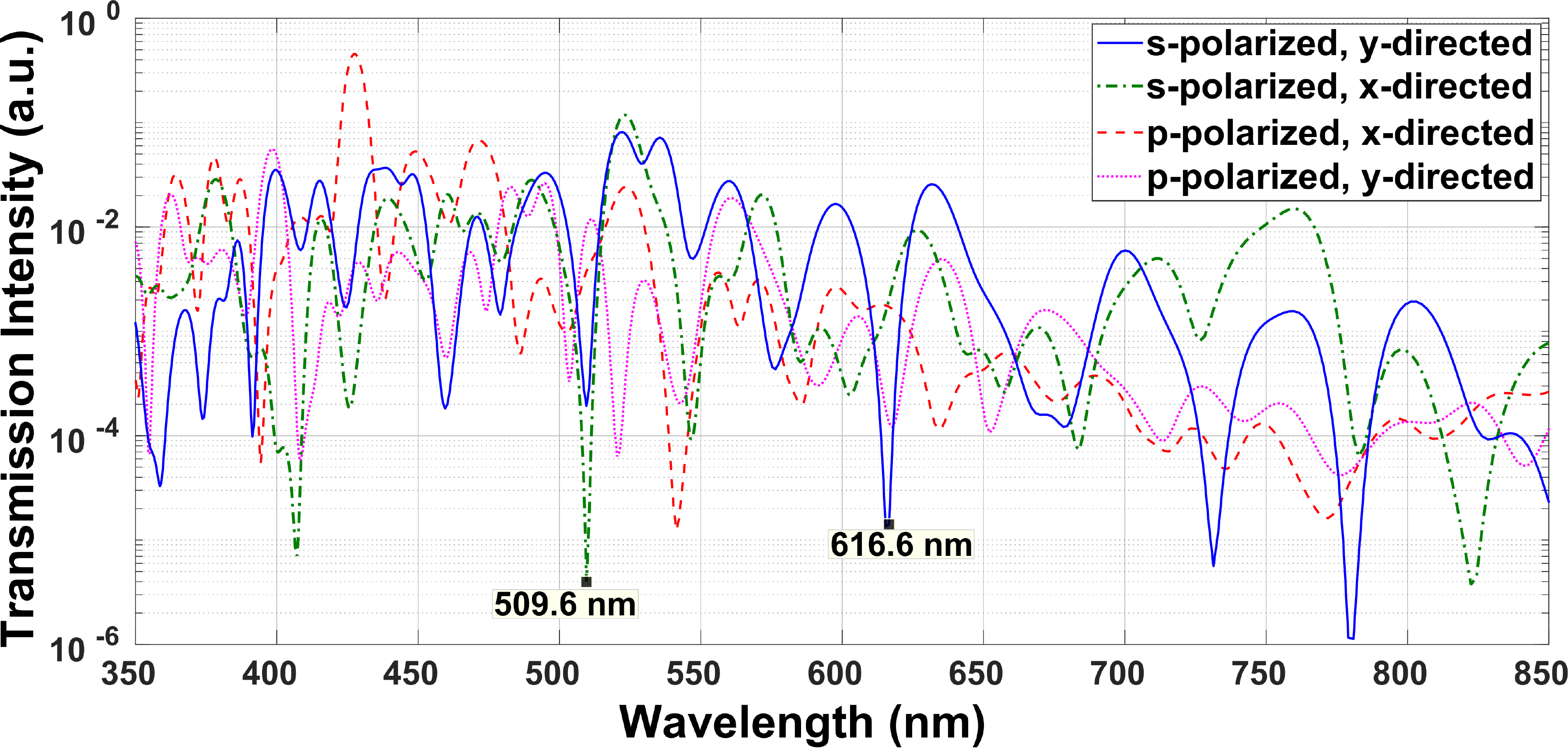}
	\caption{\label{fig:HAAOmode}Wavelength spectrum of horizontally propagated modes in PC of the proposed HPPS obtained for different directions of propagation in $xy$-plane with either s- or p-polarizations.}
\end{figure}

In order to identify the underlying cause of the emergence of such a narrow emission bandwidth, all modes propagated horizontally (along the $xz$ plane) in PC of the proposed HPPS are also calculated and shown in Fig.~\ref{fig:HAAOmode}. 
It is known that there is no possibility for p-polarized modes to be vertically propagated since the associated electric field is aligned with z-axis. 
On the other hand the electrical field is directed horizontally for s-polarized modes and thus, for each wavelength that the horizontal propagation is prevented by HPPS, a vertical reflection is possible.
Therefore, considering that the excitation input wave is aligned with y-axis, one should focus on the y-directed s-polarized propagation graph in Fig.~\ref{fig:HAAOmode}, which reveals a trough at $\lambda=616$ nm.  
In this sense, the portion of the emission band of S101 molecules that coincides with this trough is expected to reflect vertically while the rest propagates along the metal surface.
Since for S101 molecules, the excitation band overlaps partially with the emission band, the horizontally propagated modes can excite the neighboring fluorophores.
This synchronizes the transitions of the molecules and therefore, a spatial coherence is also expected. 
The visibilities obtained as results of the double-slits tests are shown in Fig.~\ref{fig:AAOFlu} (inset), which clearly shows a spatial coherence width of greater than $10\:\mu$m. 
Anyway, this coherence width is proportional to the propagation length of the horizontally propagated modes, which is determined by the imaginary part of the corresponding wave-vectors. It can state that in the near-field, the plasmonic modes are responsible for coherence while the s-polarized modes are capable to transfer this coherence to the far-field. It must also be noted that the conversion between the s- and p-polarized modes is possible due to the random orientation of the dipole moment of the molecules and internal reflections inside cavities.
It is worth noting that the wider the spectral range of the horizontally propagated modes, the larger the portion of energy absorbed by structure. This is the case for the proposed HPPS as seen in Fig.~\ref{fig:HAAOmode}, where only a small portion of modes (corresponding to the troughs) is prevented from being horizontally propagated, and thus, a large portion of the incident energy is responsible for exciting the fluorophores (see Fig.~\ref{fig:AAOFlu}). This is also among desirable features of the proposed HPPS compared to the previously proposed structures.

%
%

%


%
\end{document}